\documentclass[singlespacing]{elsart}
\usepackage{amsmath}
\usepackage{amssymb}
\usepackage{graphics}
\usepackage{graphicx}
\usepackage{epsfig}
\usepackage{dcolumn}
\usepackage{bm}
\usepackage{lineno}

\begin{document}
RD51-NOTE-2015-004
\begin{frontmatter}
\title{Design and fabrication of a data logger for atmospheric pressure, temperature and relative humidity for gas-filled detector development}
\author[label1]{S.~Sahu},
\author[label2]{M.~R.~Bhuyan},
\author[label3]{Sharmili~Rudra},
\author[label2]{S.~Biswas\corauthref{cor}},
\ead{saikat.ino@gmail.com, s.biswas@niser.ac.in, saikat.biswas@cern.ch}
\corauth[cor]{}
\author[label2]{B.~Mohanty},
\author[label1]{P.~K.~Sahu}


\address[label1]{Institute of Physics, Sachivalaya Marg, P.O: Sainik School, Bhubaneswar - 751 005, Odisha, India}
\address[label2]{School of Physical Sciences, National Institute of Science Education and Research, Jatni - 752050, India}
\address[label3]{Department of Applied Physics, CU, 92, APC Road, Kolkata-700 009, West Bengal, India}

\begin{abstract}
A novel instrument has been developed to monitor and record the ambient parameters such as temperature, atmospheric pressure and relative humidity. These parameters are very essential for understanding the characteristics such as gain of gas filled detectors like Gas Electron Multiplier (GEM) and Multi Wire Proportional Counter (MWPC). In this article the details of the design, fabrication and operation processes of the device has been presented.   
\end{abstract}
\end{frontmatter}

\section{Introduction}\label{intro}
Keeping in mind the emerging needed to build and test micro-pattern gas detector (MPGD) such as Gas Electron Multiplier (GEM) for several upcoming High-Energy Physics (HEP) experiment projects \cite{ALICE}, we have also taken up an initiative to build and characterise the performance of such detectors. Temperature (t), atmospheric pressure (p) and relative humidity (RH) monitor and recording is very important for understanding the responses of the gas filled  detectors \cite{SB}. The effective gain of the GEM varies with absolute temperature T (=t+273) in Kelvin and pressure p in atmospheric pressure as 
\begin{equation}\label{eq1}
G(T/p)={Ae^{B\frac{T}{p}}}
\end{equation}
where, A and B are fit parameters, determined by fitting the curve of measured gain and $\frac{T}{p}$ by the exponential function \cite{MCA03}. Here t is expressed in $^{\circ}\mathrm{C}$.

The data logger designed here measures temperature, RH and atmospheric pressure. The system consists of two parts one is hardware based data acquisition system (DAQ) and other one is the corresponding DAQ Software. The hardware DAQ system is designed with a micro-controller based system with a 16$\times$4 line Alphanumeric LCD display unit. The display unit can update with a minimum of 2-3 sec interval. The interval can be made longer to about a few minutes. The hardware DAQ system has one external power port for 9~V DC and one RS232 communication port to interface to a PC for interfacing with data logger software. The DAQ system transmits the measured information with following units, temperature in $^{\circ}\mathrm{C}$, RH in \% and atmospheric pressure in mbar. The details of the fabrication and operation of the data logger is presented.

The paper is organised as follows. In the next section we discuss the data logger in terms of the circuit diagram, micro controller algorithm and LabView programming. The section~\ref{sw} discusses the operation of the data logger. In section~\ref{res} we present some typical measurements of temperature, pressure and relative humidity. Finally in section~\ref{con} we summarise and present a brief advantages of such a data logger.

\section{Description of the data logger}\label{construct}

\begin{figure}[htb!]
\begin{center}
\includegraphics[scale=0.5]{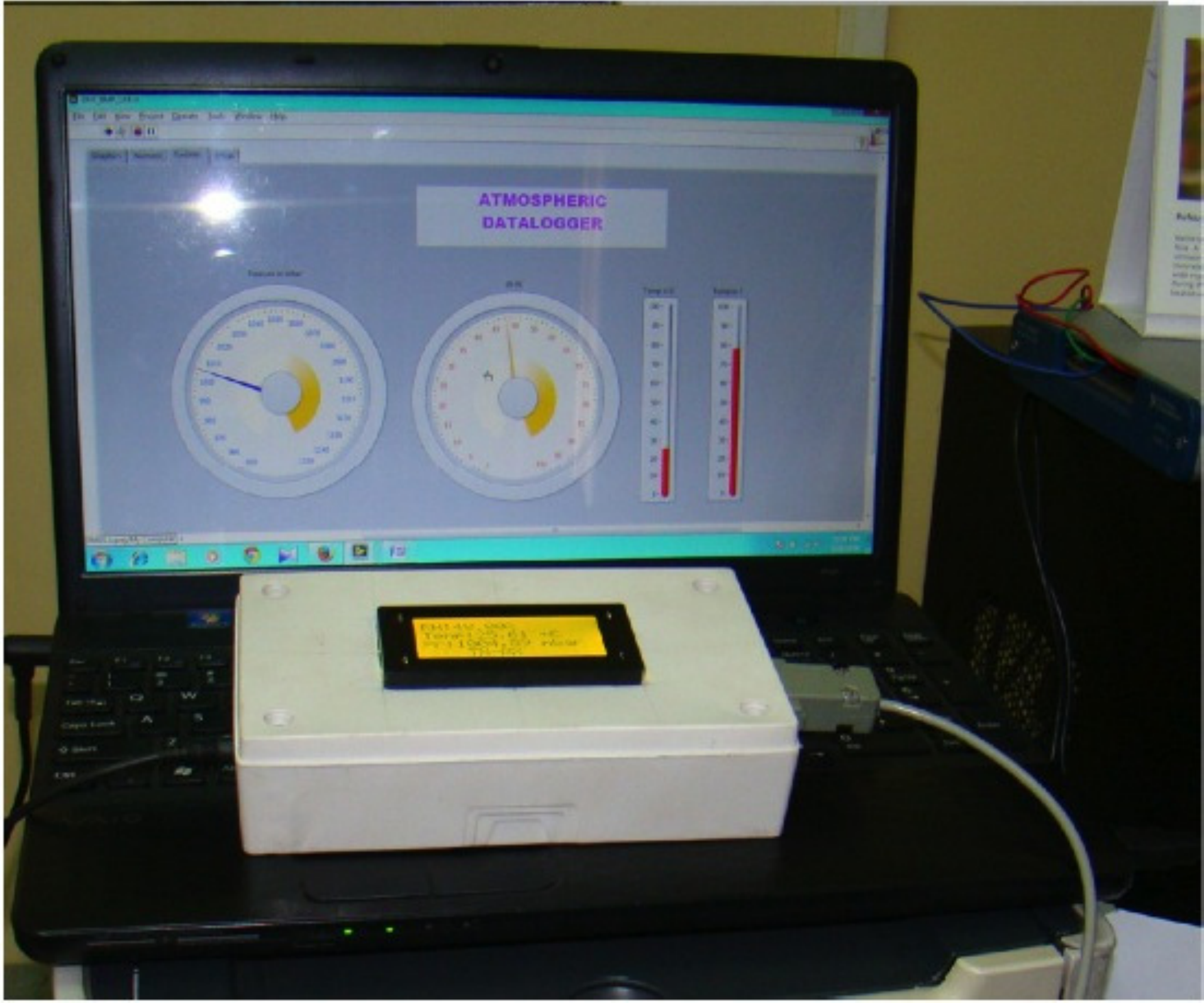}
\caption{\label{hw}DAQ hardware. The white box contains the electrical circuits including all sensors. In the computer screen from the left the circular gadgets show the pressure in mbar, relative humidity in \% and the vertical linear gadgets show the temperature in $^{\circ}\mathrm{C}$ and $^{\circ}\mathrm{F}$ respectively.}\label{hw}
\end{center}
\end{figure}
The data logger hardware unit is shown in Figure~\ref{hw} and the display screen is shown in Figure~\ref{display}.
For the data logger the power supply specifications are as follows: Input Voltage is 230 Volt at a frequency of  50-60Hz; Output Voltage is 9-12~V DC and Output Current is 500-750~mA.

\begin{figure}[htb!]
\begin{center}
\includegraphics[scale=0.4]{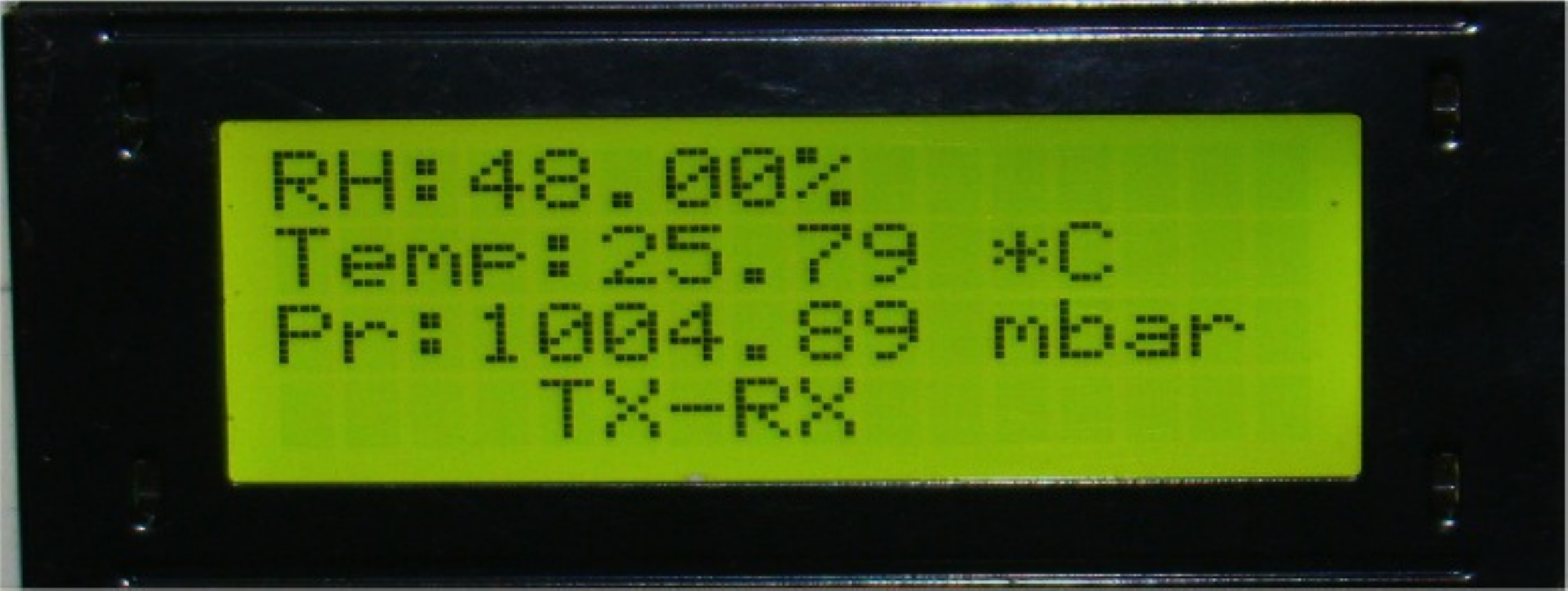}
\caption{\label{display}The LCD Display unit is consisting of 16$\times$4 line Alphanumeric Display. The Display unit is Backlighted for better visibility.}\label{display}
\end{center}
\end{figure}
\begin{figure}[htb!]
\begin{center}
\includegraphics[scale=0.6]{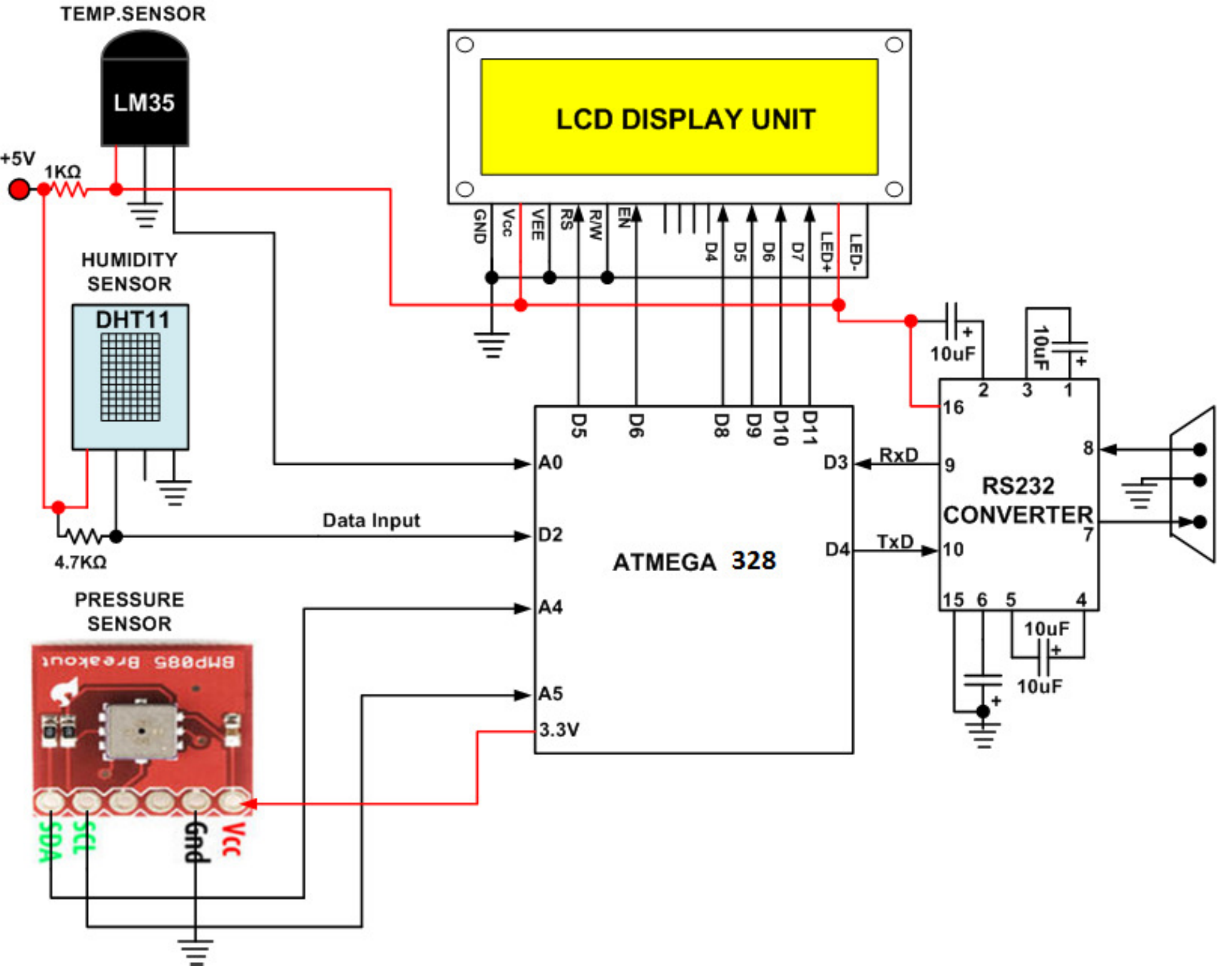}
\caption{\label{circuit}The circuit diagram of the data logger.}\label{circuit}
\end{center}
\end{figure}
\begin{figure}[htb!]
\begin{center}
\includegraphics[scale=0.35]{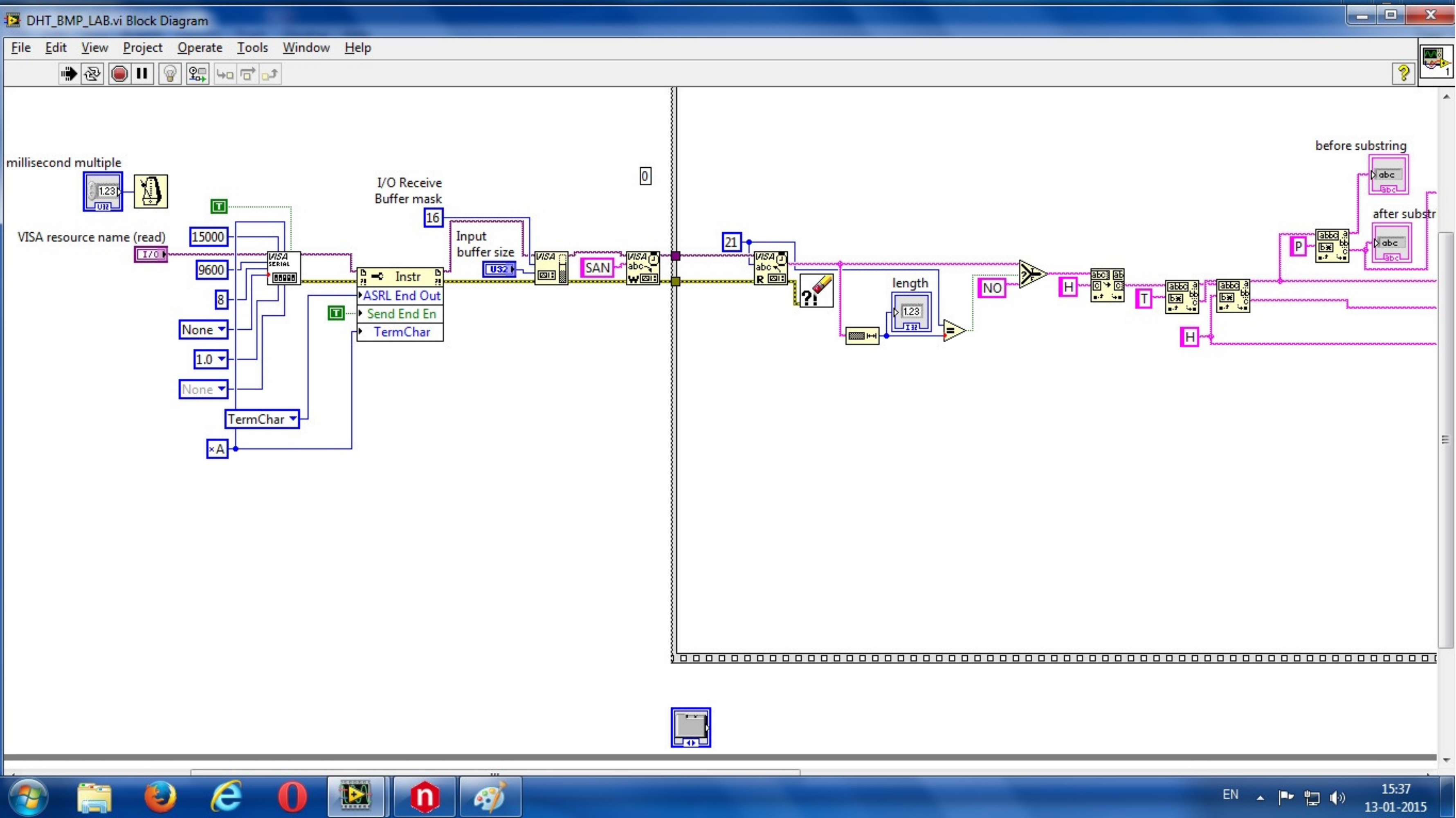}
\includegraphics[scale=0.35]{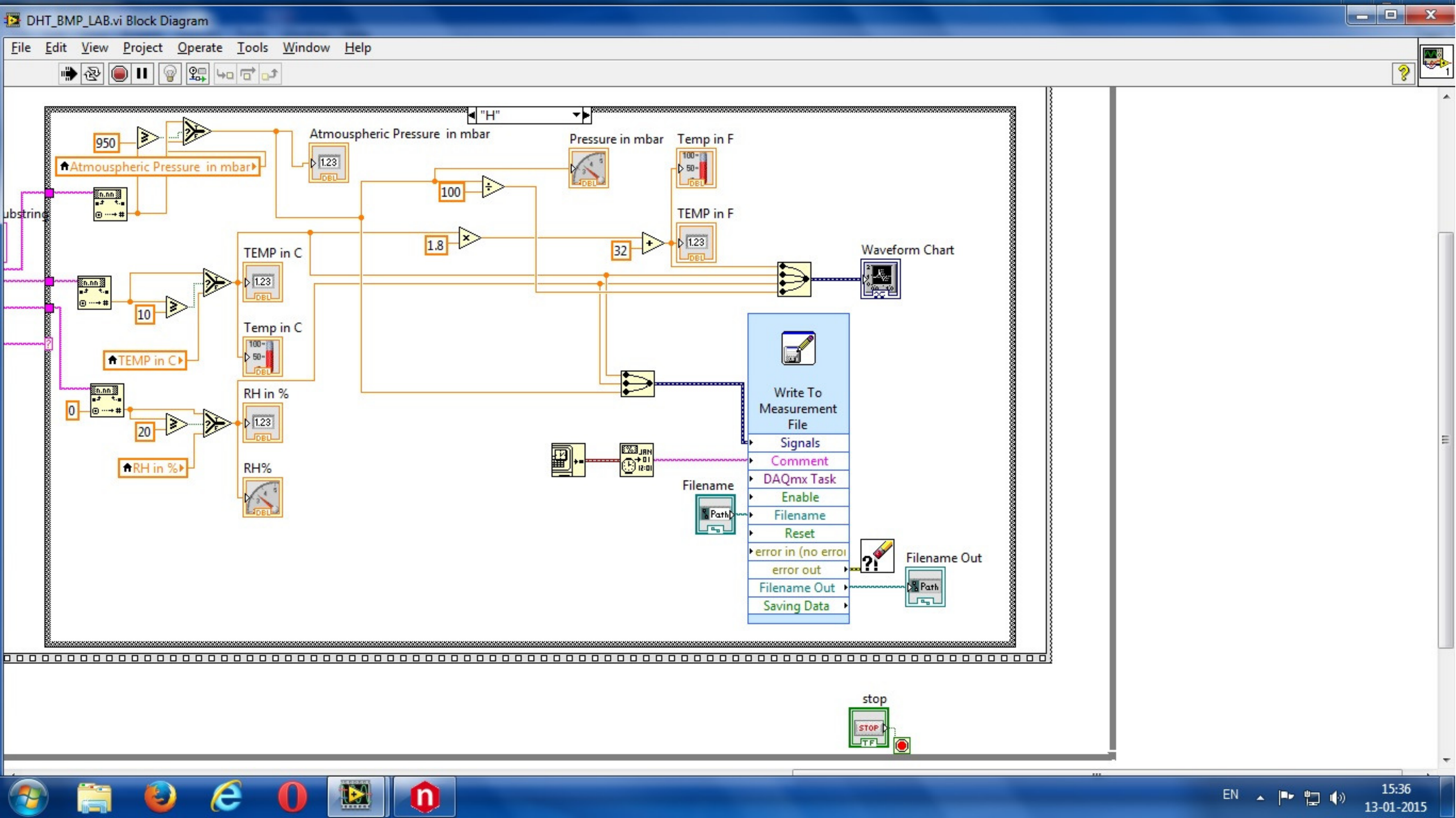}
\caption{\label{block}The LabView block diagram (in two parts).}\label{block}
\end{center}
\end{figure}

The detail circuit diagram of the data logger is shown in Figure~\ref{circuit}. In this data logger LM35 temperature sensor, BMP085 pressure sensor and DHT11 humidity sensor have been used. The LabView block diagram are shown in two parts in Figure~\ref{block}. The algorithms for the Micro controller and for the LabView programming are the following:

{\bf Algorithms of Micro Controller:}

1.	Initialize the Liquid Crystal Display, Atmega328P Controller and memory array  and Humidity Sensor DHT11, Pressure sensor BMP085.
 
2.	The DHT11 is initialized for single wire two-way communications and BMP085 is initialized for I2C two-wire communication.

3.	The LM35 precession temperature sensor is interfaced to the inbuilt 10bit ADC of the controller.

4.	The analogue voltage corresponding to the temperature are sampled with a 20ms sampling interval. The controller stores 50 such samples in the specified memory array and removes the high frequency noise by averaging method.

5.	Read the BMP085 pressure sensor using I2C  communication method.

6.	Read the DHT11 Humidity sensor and using single wire bi-directional communication method.

7.	Display the temperature in degree Centigrade and atmospheric pressure in mbar in the 16x4line Liquid Crystal Display unit interfaced to the controller.
 
8.	The controller check for the command from the  computer to send data, if received then send all the information to the PC in a string format and go to step no. 4, else the control goes to step no. 4.

{\bf Algorithm of LabView Programming:}

1.	The LabView based program is designed to read the data from the micro-controller by sending a Read Command and the interval of reading  can be set by the user.
 
2.	The string received by the PC is defragmented to separate the data and initial markers.

3.	The data is further processed to convert in different scale and display graphically on a dynamic window.

4.	The temperature, relative humidity and atmospheric pressure are displayed in numerical and gadgets also in different tabs.

5.	These numeric data are stored in a spreadsheet  file continuously for logging and the same data is used for correction and calculation purpose. After completing this procedure  the control goes to step 1 to repeat the procedure. 

The flow chart for the Micro controller and LabView programming are shown in Figure~\ref{algorithm_MC} and~\ref{algorithm_LabView} respectively.

\begin{figure}[htb!]
\begin{center}
\includegraphics[scale=0.5]{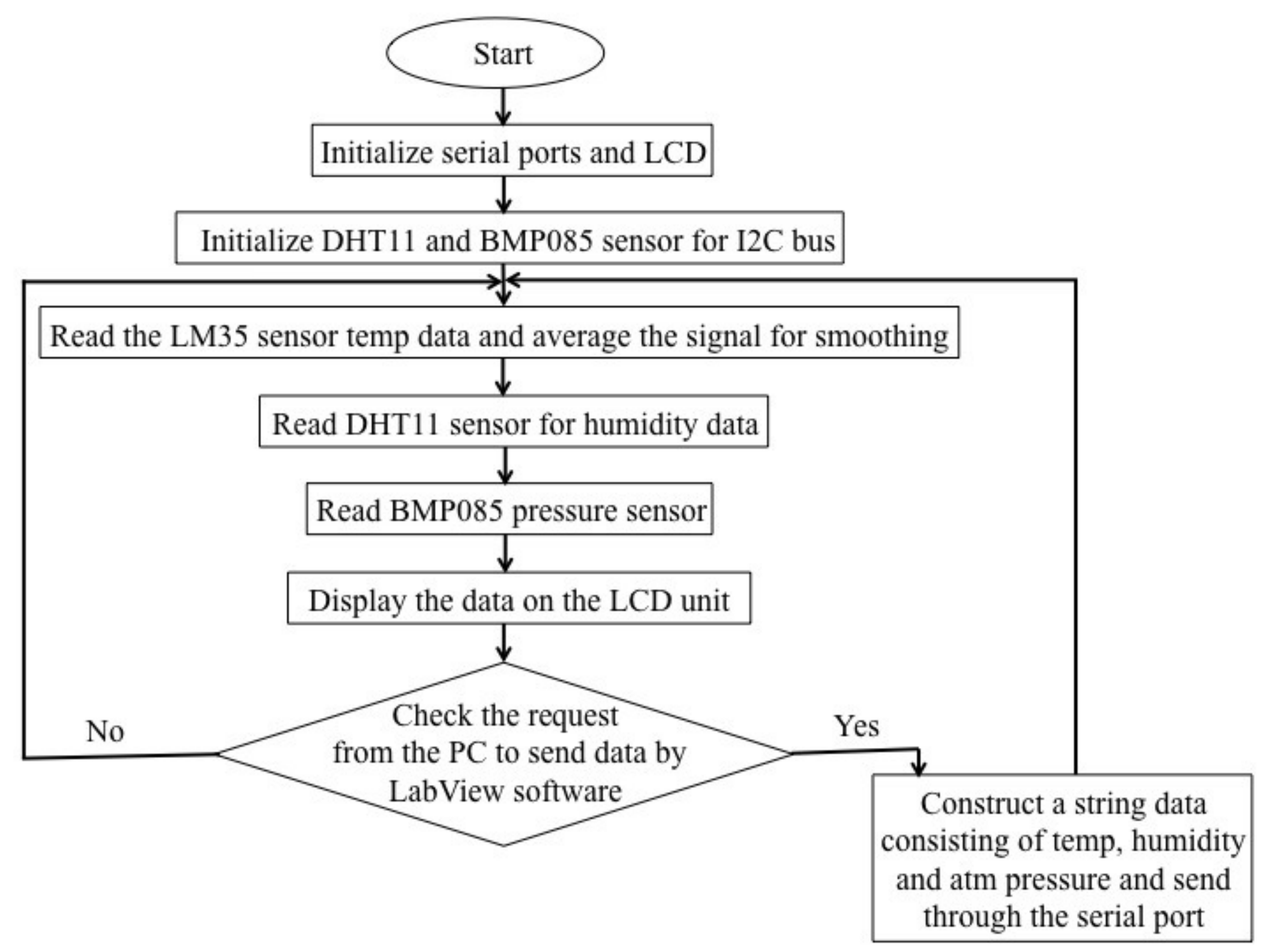}
\caption{\label{algorithm_MC}The flow chart of the Micro controller.}\label{algorithm_MC}
\end{center}
\end{figure}
\begin{figure}[htb!]
\begin{center}
\includegraphics[scale=0.5]{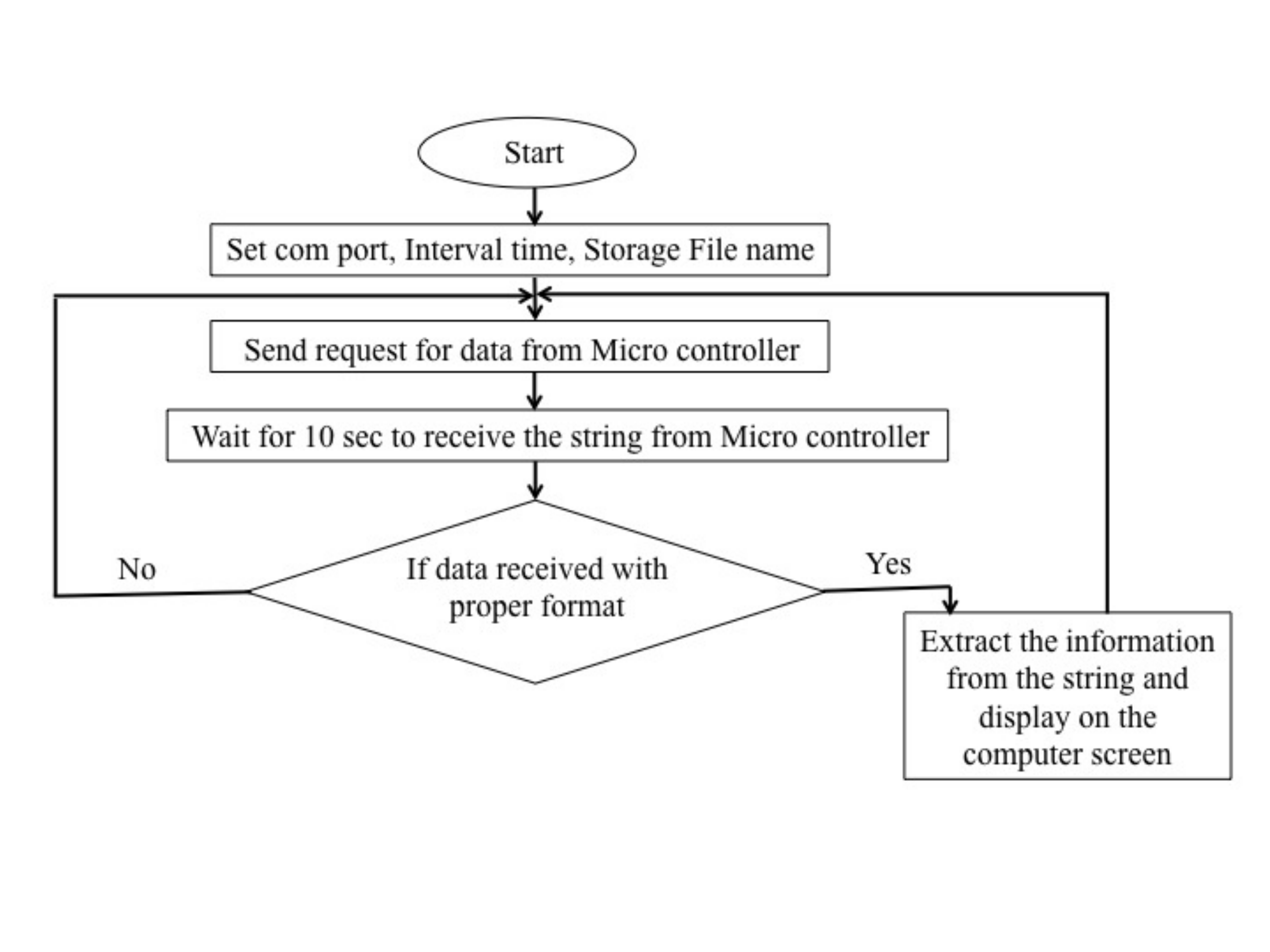}
\caption{\label{algorithm_LabView}The flow chart of the LabView programming.}\label{algorithm_LabView}
\end{center}
\end{figure}

The parameters for the data logger are as follows:

1. Temperature: Measurement of temperature can be done with 0.25$^{\circ}\mathrm{C}$ assured accuracy. The rated full range of measurement is from 0$^{\circ}\mathrm{C}$ to 150$^{\circ}\mathrm{C}$. It has very low self-heating with resolution $\sim$ 0.01$^{\circ}\mathrm{C}$.

2. RH: Measurement of RH can be done with 1\% resolution and with 4\% accuracy. The stability varies $\pm$1\%RH/Year and the hysteresis is 1\% RH.

3. Pressure: Measurement of pressure can be done with a range: 300-1100~hPa (+9000~m to -500~m above sea level). Low noise: 0.06~hPa (0.5~m) in ultra low power mode. 0.03~hPa (0.25~m) ultra high resolution mode and less than 0.1~m is possible with software averaging algorithm.

\section{Data logger software and it's operation}\label{sw}

The data logger software is designed and developed with LabView platform of National Instrument, USA \cite{NI}. The software developed with an idea to provide maximum user friendly operation. There are multiple tabs in the application named as follows:

\textbf{Setup:} Basically this tab is used for the user to configure the DAQ system as per the requirement. The set-up tab is shown in Figure~\ref{setup}.

\begin{figure}[htb!]
\begin{center}
\includegraphics[scale=0.5]{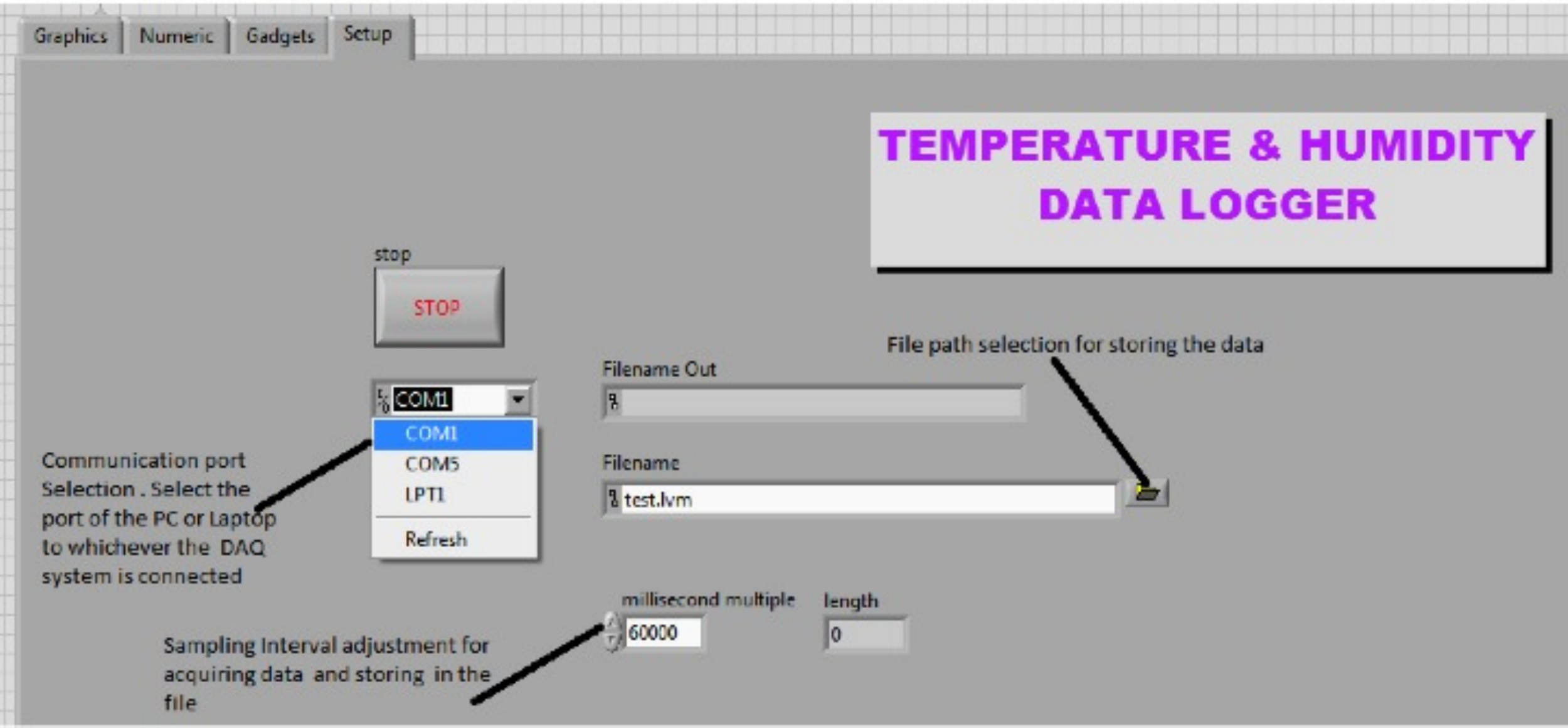}
\caption{\label{setup}The set-up tab.}\label{setup}
\end{center}
\end{figure}
\begin{figure}[htb!]
\begin{center}
\includegraphics[scale=0.35]{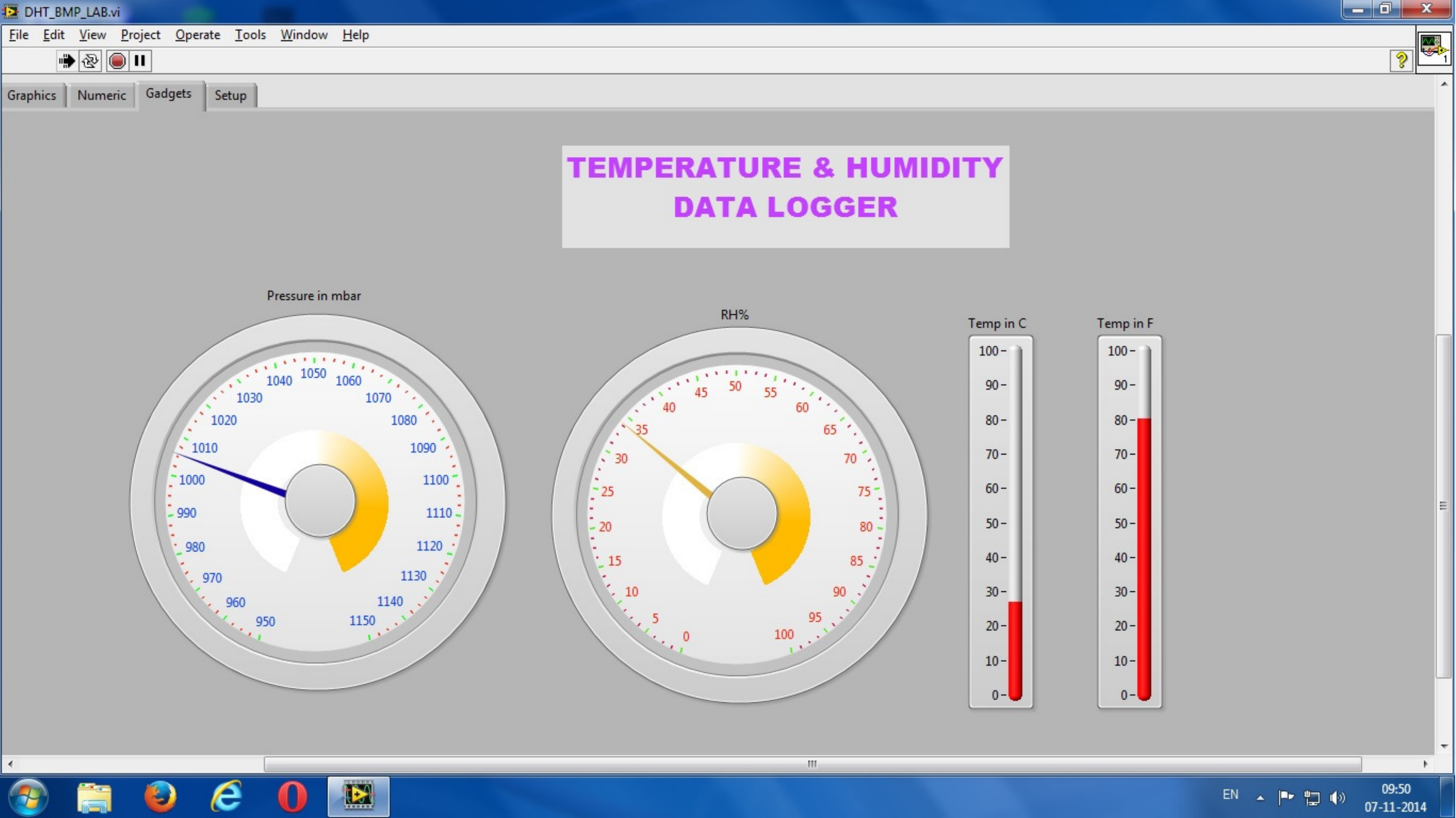}
\caption{\label{gadget}The gadget tab.}\label{gadget}
\end{center}
\end{figure}
\begin{figure}[htb!]
\begin{center}
\includegraphics[scale=0.35]{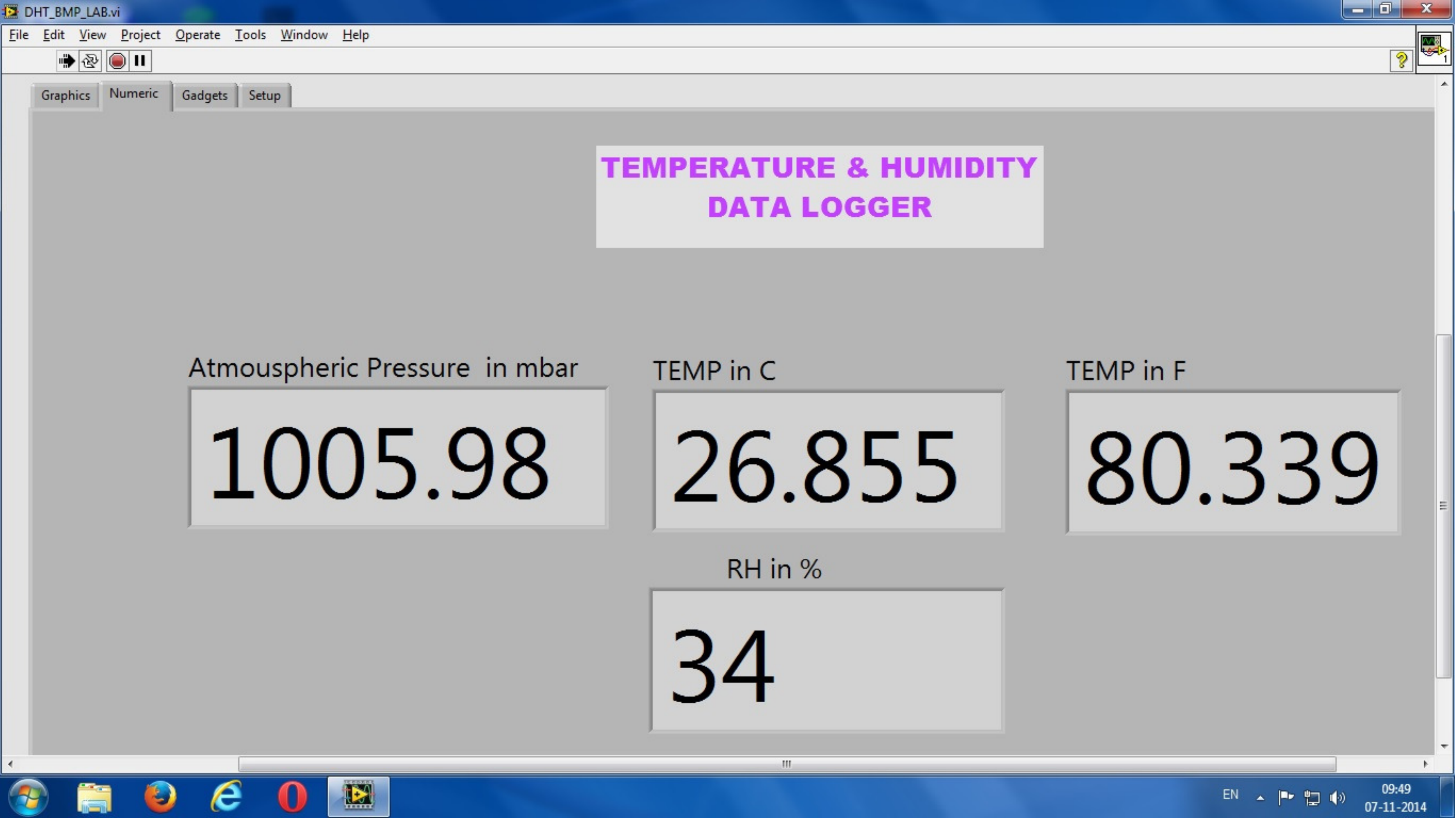}
\caption{\label{numeric}The numeric tab.}\label{numeric}
\end{center}
\end{figure}
\begin{figure}[htb!]
\begin{center}
\includegraphics[scale=0.35]{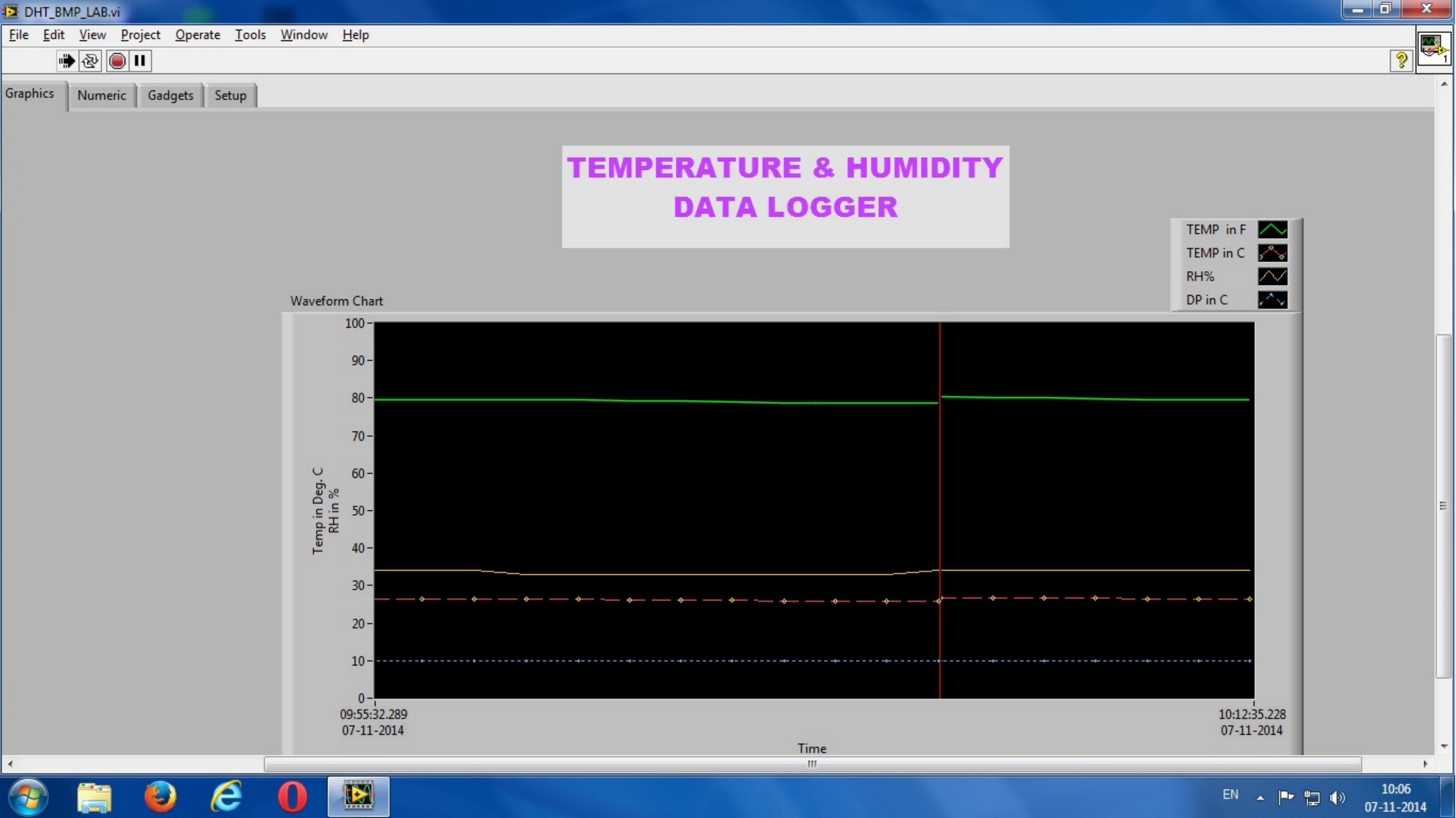}
\caption{\label{graphic}The graphic tab.}\label{graphic}
\end{center}
\end{figure}

The software operations are as follows:

1. The communication port (COM PORT) needs to be selected properly. 

2. The sampling interval in msec needs to be entered by the user.

3.The file path needs to be entered for data storing.

\textbf{Gadget:} In this tab the parameters are displayed in Gadgets for example, pressure and RH in two separate Dial Gauge. The temperature in $^{\circ}\mathrm{C}$ and $^{\circ}\mathrm{F}$ is displayed in a thermometer. The gadget tab is shown in Figure~\ref{gadget}.

\textbf{Numeric:} In Numeric Tab all the parameters are displayed in numerical windows for example, pressure in mbar , RH in \% , temperature in $^{\circ}\mathrm{C}$ and $^{\circ}\mathrm{F}$ in decimal format. The numeric tab is shown in Figure~\ref{numeric}.

\textbf{Graphic:} In this Tab all the parameters temperature in $^{\circ}\mathrm{C}$ or in $^{\circ}\mathrm{F}$, RH in \% and pressure at in mbar/100 are displayed on a line graph window with respect to time axis. The graphic tab is shown in Figure~\ref{graphic} \cite{SS, SBRD51}.

\section{Result}\label{res}
The temperature, pressure and RH measured in 24 hours has been plotted as a function of time in Figure~\ref{result}. The variation of these three ambient parameters during the day and night is observed and recorded in our laboratory situated in Bhubaneswar, India.\\

\begin{figure}[htb!]
\begin{center}
\includegraphics[scale=0.44]{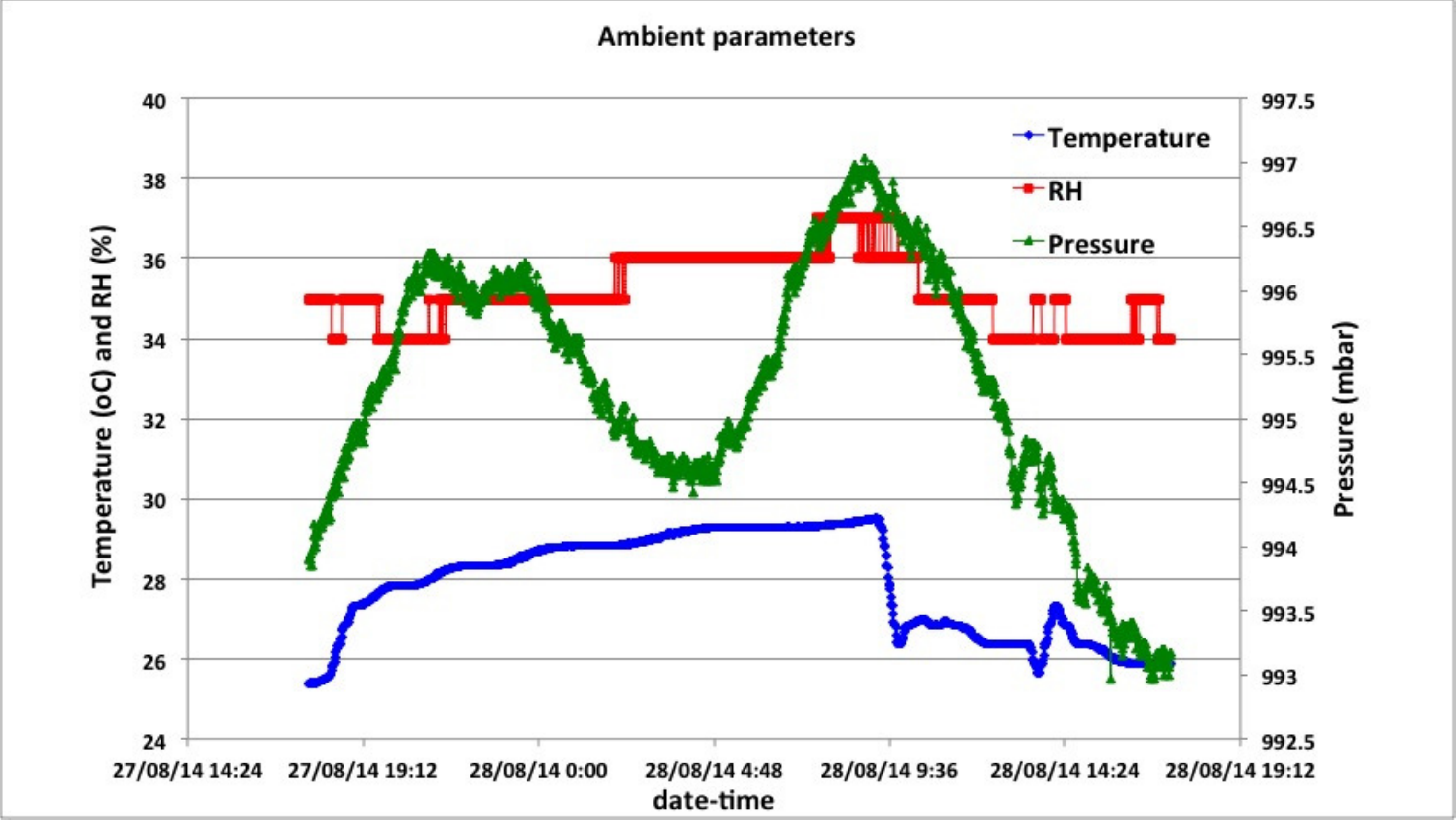}
\caption{\label{result}Temperature, pressure and RH measured in 24 hours as a function of date-time. The horizontal axis shows the date and time whereas the left vertical axis shows the temperature in $^{\circ}\mathrm{C}$ and RH in \%. The right vertical axis shows the atmospheric pressure in mbar.}\label{result}
\end{center}
\end{figure}
\begin{figure}[htb!]
\begin{center}
\includegraphics[scale=0.4]{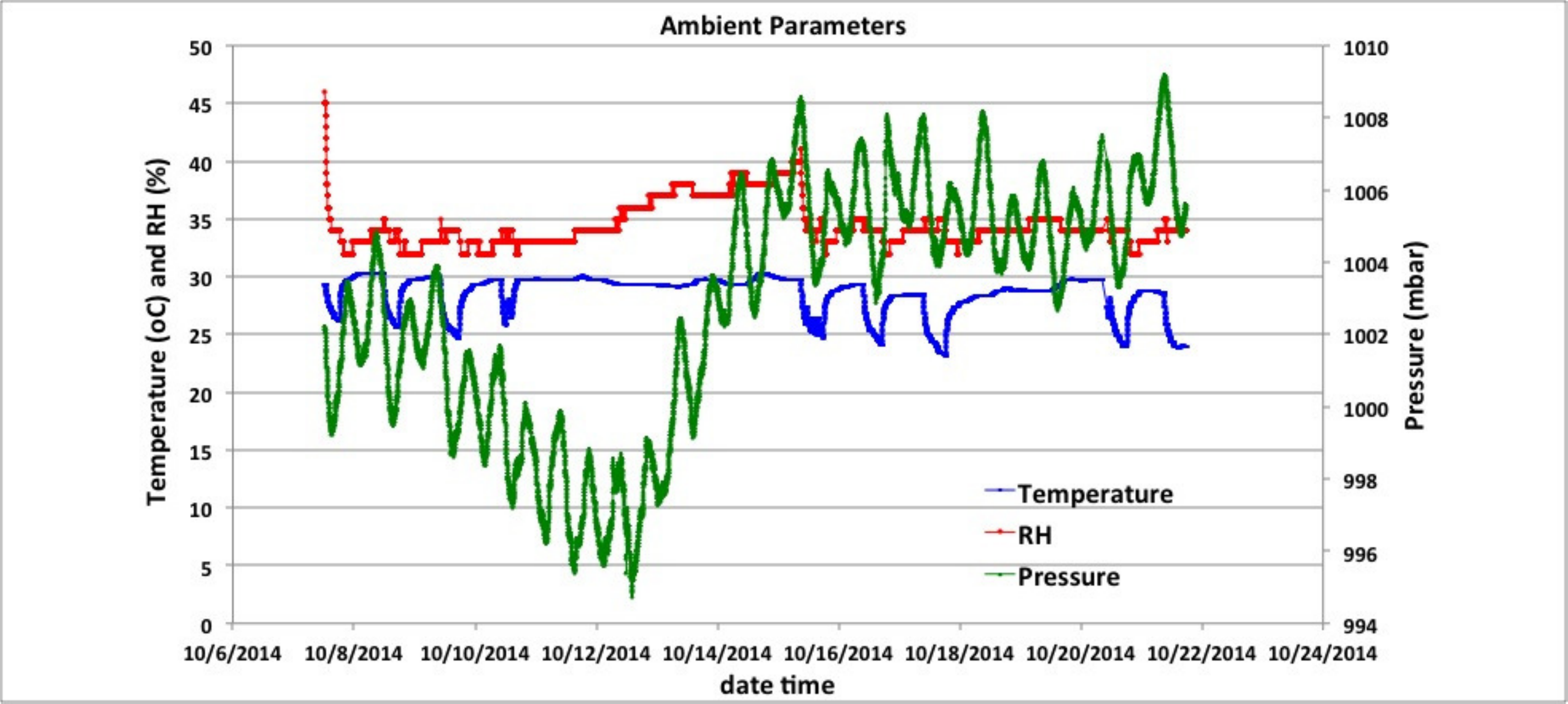}
\caption{\label{result_long}Temperature, pressure and RH measured continuously for 15 days, as a function of date-time.}\label{result_long}
\end{center}
\end{figure}

These three ambient parameters measured continuously for 15 days in an interval of 60 seconds has been plotted as a function of date-time in Figure~\ref{result_long}. A periodic variation in pressure is observed everyday. This particular data set has a special significance. During 12-13 October 2014 there was a high depression in the environment along with a storm (cyclone Hudhud) in the coastal region of Andhra Pradesh and Odisha (area of our institute where the measurements are taken). So the signature of cyclone Hudhud is captured by the data logger. This data logger shows a very low pressure compared to the average normal value, particularly on those two days.

\section{Conclusions and outlooks}\label{con}
A data logger to monitor and record the atmospheric parameters such as temperature, relative humidity and pressure in a laboratory set-up for gas filled detector has been developed. With this data logger continuous recording of temperature, atmospheric pressure, relative humidity and the time stamp can be done with a programmable sampling interval. This instrument is very cheap and the data of temperature, atmospheric pressure and relative humidity, measured by this instrument are very essential to understand the gain of a gas filled detector such as GEM, MWPC.

\section{Acknowledgements}
We are thankful to Prof. T. K. Chandrashekar, Secretary, SERB, Prof. V. Chandrasekhar, Director, NISER and Prof. Sudhakar Panda, Director, IoP for their support in this work. S. Biswas acknowledges the support of DST-SERB Ramanujan Fellowship (D.O. No. SR/S2/RJN-02/2012). This work is also funded through DST project no. SB/S2/HEP-022/2013. XII$^{th}$ Plan DAE project titled Experimental High Energy Physics Programme at NISER-ALICE is also acknowledged.


\noindent
\begin{thebibliography}{50}

\bibitem{ALICE} Technical Design Report for the Upgrade of the ALICE Time Projection Chamber, ALICE-TDR-016, CERN-LHCC-2013-020, March 3 2014.


\bibitem{SB} S. Biswas et al.,
\emph{Development of a GEM based detector for the CBM Muon Chamber (MUCH)},
\emph{2013 JINST 8 C12002} doi:10.1088/1748-0221/8/12/C12002.

\bibitem{MCA03} M.C. Altunbas et al.,
\emph{Aging measurements with the Gas Electron Multiplier (GEM)},
\emph{Nuclear Instruments and Methods in Physics Research Section A} \textbf{515} (2003) 249.

\bibitem{NI} http://www.ni.com/labview/whatsnew/.

\bibitem{SS} S. Sahu et al., Proceedings of the DAE Symp. on Nucl. Phys. 59 (2014) 876.

\bibitem{SBRD51} S. Biswas et al., RD51 mini week, CERN (2014), "Activities on GEM detector development at NISER-IoP, India", https://indico.cern.ch/event/356113/ session/3/contribution/8/material/slides/0.pdf.  

\end{thebibliography}
\end{document}